\documentclass[prl]{revtex4-1}
\usepackage[T1]{fontenc}
\usepackage[latin9]{inputenc}
\usepackage{amsmath}
\usepackage{subscript}
\usepackage[unicode=true,pdfusetitle,
 bookmarks=true,bookmarksnumbered=false,bookmarksopen=false,
 breaklinks=false,pdfborder={0 0 1},backref=false,colorlinks=false]
 {hyperref}

\makeatletter

\pdfpageheight\paperheight
\pdfpagewidth\paperwidth

\@ifundefined{textcolor}{}
{%
 \definecolor{BLACK}{gray}{0}
 \definecolor{WHITE}{gray}{1}
 \definecolor{RED}{rgb}{1,0,0}
 \definecolor{GREEN}{rgb}{0,1,0}
 \definecolor{BLUE}{rgb}{0,0,1}
 \definecolor{CYAN}{cmyk}{1,0,0,0}
 \definecolor{MAGENTA}{cmyk}{0,1,0,0}
 \definecolor{YELLOW}{cmyk}{0,0,1,0}
}

\makeatother

\begin{document}

\title{Mapping pure gravity to strings in three-dimensional anti-de Sitter geometry}

\author{Bo Sundborg}

\affiliation{The Oskar Klein Centre}

\address{Department of Physics, Stockholm University, SE-106 91 Stockholm,
Sweden}
\begin{abstract}
Strings propagating in three-dimensional anti-de Sitter space with a background antisymmetric
tensor field are well understood, even at the quantum level. Pure three-dimensional gravity with a negative cosmological constant is potentially important because of the existence of black hole solutions and an asymptotic conformal symmetry, but it is mysterious and surprisingly resistant to analysis. In this letter, the two theories are related by a map
on the classical level. The map is obtained by gauge fixing
the string completely, like in a light cone gauge, and comparing the
resulting constrained theory with the boundary theory obtained from
gravity by imposing the appropriate asymptotic boundary conditions.
The two theories are formally related as different gauge fixings of
the same gauge theory.
\end{abstract}
\maketitle
Pure, three-dimensional gravity with a negative cosmological constant
is a perfect laboratory for subtle issues in quantum gravity: It has good UV properties \citep{Witten1988},
black holes with horizons of varying length \citep{BanadosTeitelboimZanelli1992,BanadosHenneauxTeitelboimZanelli1993},
and an infinite dimensional conformal symmetry \citep{BrownHenneaux1986}.
There are no local degrees of freedom, and and continuum degrees of
freedom are only associated with the boundaries of spacetime. These
assets notwithstanding, it is still unknown exactly how to get a sensible
spectrum for this theory \citep{Carlip2009,MaloneyWitten2010}. The
local geometry of all the classical solutions is three-dimensional
anti-de Sitter space (AdS\textsubscript{3}). In contrast to the gravity
case, strings in AdS\textsubscript{3} are well under control, given
the correct additional antisymmetric tensor background to guarantee
conformal invariance and consistent string propagation. Although there
were initial issues with potential negative norm states \citep{BalogORaifeartaighForgacsWipf1989},
ghost free and modular invariant truncations of the spectrum indicated
that there was a resolution \citep{Hwang1991,HenningsonHwangRobertsSundborg1991,Hwang1992},
and finally a coherent picture emerged \citep{MaldacenaOoguri2001,MaldacenaOoguriSon2001,MaldacenaOoguri2002}.
The goal of the present work is to relate gravity directly to the known string theory, in order to make string theory
 results applicable to AdS\textsubscript{3} gravity. The space-time anti-de Sitter symmetry of both theories is a product
of special linear transformations, SL(2,R)$\times$SL(2,R). Furthermore,
the fact that a conformal SL(2,R) Wess-Zumino-Witten (WZW) model figures
in both AdS\textsubscript{3} gravity and string theory is a naive
hint that there may be a relation between them, although the role
of the WZW model is quite different in the two theories. This letter
bridges the differences and constructs a local map between the
classical theories in the WZW framework.

\paragraph*{AdS\textsubscript{3} gravity.}

In the Chern-Simons formulation \citep{AchucarroTownsend1986,Witten1988}
of AdS\textsubscript{3} gravity the vielbein $e$ is written in terms
of two SL(2,R) connections, $e=A-\bar{A}$, and the equations of motion
amount to their flatness conditions $F=\bar{F}=0$, implying that
spacetime is locally AdS\textsubscript{3}. We consider this theory
 on $\mathcal{A\times R}$, an annulus times the real line,
so the boundary consists of two cylinders, on which the degrees of
freedom of the theory live \citep{MooreSeiberg1989,ElitzurMooreSchwimmerSeiberg1989}.
We will actually only focus on the outer boundary, but without the presence of another boundary the 
classical solutions would be unnecessarily
constrained. The spacetime geometry is  asymptotically AdS\textsubscript{3}, 
implying the boundary conditions 
studied in \citep{CoussaertHenneauxDriel1995} based on the earlier
discussion by Brown and Henneaux \citep{BrownHenneaux1986}. The boundary
conditions lead to a non-chiral WZW model, and to its Hamiltonian
reduction to Liouville theory (see also e.g. \citep{ForgacsWipfBalogFeherORaifeartaigh1989}).

We write the SL(2,R) WZNW model in terms of the SL(2,R) valued field
$G(\tau+\sigma,$$\tau-\sigma)=G(\xi,\bar{\xi})$. (The circle coordinate
$\sigma$ will be important below, where periodicity is discussed.)
The equations of motion are current conservation equations
\begin{equation}
\bar{\partial}J\left[\lambda\right]=0,\quad\partial\bar{J}\left[\lambda\right]=0,\label{eq:eom}
\end{equation}

\noindent with $\partial=\partial_{\xi}$ and $\bar{\partial}=\partial_{\bar{\xi}}$,
in terms of the currents
\begin{equation}
\begin{array}{c}
J\left[\lambda\right]=\kappa\mathrm{\, Tr}\left\{ \lambda\cdot\partial G\: G^{-1}\right\} =\kappa\mathrm{\, Tr}\left\{ \lambda\cdot g'g^{-1}\right\} ,\\
\bar{J}\left[\lambda\right]=-\kappa\,\mathrm{Tr}\left\{ \lambda\cdot G^{-1}\bar{\partial}G\right\} =-\kappa\,\mathrm{Tr}\left\{ \lambda\cdot\bar{g}^{-1}\bar{g}'\right\} ,
\end{array}\label{eq:currents}
\end{equation}

\noindent where $\lambda$ are SL(2,R) Lie algebra elements, and a
general solution is
\begin{equation}
G(\xi,\bar{\xi})=g(\xi)\bar{g}(\bar{\xi)}.\label{eq:solution}
\end{equation}

Next, we need to understand how these WZW solutions are related to
gravity and its asymptotic boundary conditions. As described in \citep{ForgacsWipfBalogFeherORaifeartaigh1989}
the Gauss decomposition $G=ABC$ with 
\begin{equation}
A=\begin{pmatrix}1 & X\\
0 & 1
\end{pmatrix}=\exp\left(XE_{+}\right),\quad C=\begin{pmatrix}1 & 0\\
Y & 1
\end{pmatrix}=\exp\left(YE_{-}\right),\label{eq:gauss}
\end{equation}

\noindent and
\begin{equation}
B=\begin{pmatrix}\exp\left(\frac{1}{2}\Phi\right) & 0\\
0 & \exp\left(-\frac{1}{2}\Phi\right)
\end{pmatrix}=\exp\left(\frac{1}{2}\Phi H\right),
\end{equation}

\noindent parametrizes SL(2,R) in a way that prepares for a reduction
to Liouville theory. Supposing that left and right movers can be decomposed
analogously, $g=abc$ and $\bar{g}=\bar{a}\bar{b}\bar{c}$ and one
can check that
\begin{equation}
\exp\left(\Phi\right)=\frac{\exp\left(\phi\right)\exp\left(\bar{\phi}\right)}{\left(1+y\bar{x}\right)^{2}},
\end{equation}
 where we have introduced the notation that lower case letters denote
the left- and right-moving fields, distinguished by presence or absence
of a bar. The constraints 
\begin{equation}
y'(\xi)=\exp\left(\phi(\xi)\right),\quad\bar{x}'(\bar{\xi})=\exp\left(\bar{\phi}(\bar{\xi})\right),\label{eq:xyphi}
\end{equation}

\noindent are compatible with the equations of motion (\ref{eq:eom})
and lead to 
\begin{equation}
\exp\left(\Phi(\xi,\bar{\xi})\right)=\frac{y'(\xi)\bar{x}'(\bar{\xi})}{\left(1+y(\xi)\bar{x}(\bar{\xi)}\right)^{2}}
\end{equation}

\noindent which is a general solution of the Liouville equation. Indeed,
equations (\ref{eq:xyphi}) can be obtained from the constraints
\begin{equation}
J\left[E_{+}\right]=\kappa,\quad\bar{J}\left[E_{-}\right]=-\kappa,\label{eq:constraints}
\end{equation}

\noindent which are actually two of the constraints on the non-chiral
WZW model that were derived \citep{CoussaertHenneauxDriel1995} from
the asymptotic AdS conditions. The remaining two are
\begin{equation}
J\left[H\right]=0,\quad\bar{J}\left[H\right]=0,\label{eq:time-like}
\end{equation}

\noindent which fix the ``gauge freedom'' $G\rightarrow a(x(\xi))\, G\,\bar{c}(\bar{y}(\bar{\xi}))$,
giving a local one-one correspondence between Liouville solutions
and AdS\textsubscript{3} gravity solutions via the constrained WZNW
model. A completely gauge fixed 3d metric for each Liouville stress
tensor has been determined by Ba\~nados \citep{Banados1998}.

\paragraph*{AdS\textsubscript{3} conformal strings. }

The conformal non-chiral SL(2,R) WZW model can be interpreted as a
string propagating in the SL(2,R) group manifold, provided the Virasoro
constraints obtained upon fixing the conformal gauge are maintained.
The SL(2,R) geometry is locally AdS\textsubscript{3} but a background
anti-symmetric tensor field is also needed for conformal invariance,
and it is included in the WZW model \citep{BalogORaifeartaighForgacsWipf1989}.
The Virasoro constraints are just the vanishing of the energy-momentum
tensor due to diffeomorphism invariance of the string:
\begin{equation}
\begin{array}{c}
T=\frac{1}{2\kappa}\left[\frac{1}{2}J[H]^{2}+2J[E_{+}]J[E_{-}]\right]=0,\\
\bar{T}=\frac{1}{2\kappa}\left[\frac{1}{2}\bar{J}[H]^{2}+2\bar{J}[E_{+}]\bar{J}[E_{-}]\right]=0.
\end{array}\label{eq:Virasoro}
\end{equation}

The periodicity of string solutions is crucial. For their description
a parametrization used by Maldacena and Ooguri \citep{MaldacenaOoguri2001}
proves more illuminating than the Gauss decomposition. In particular,
the SL(2,R) winding numbers \citep{HenningsonHwangRobertsSundborg1991}
are changed by simple ``spectral flow'' transformations. With $\sigma_{i}$
the standard Pauli matrices, the flow transforms a classical WZW solution
to a new solution: 
\begin{equation}
G(\xi,\bar{\xi})\rightarrow\exp\left(\frac{i}{2}w\xi\sigma_{2}\right)G(\xi,\bar{\xi})\exp\left(\frac{i}{2}\bar{w}\bar{\xi}\sigma_{2}\right).\label{eq:flow}
\end{equation}

\noindent In the parametrization
\begin{equation}
G(\xi,\bar{\xi})=\exp\left(iu\sigma_{2}\right)\exp\left(\rho\sigma_{3}\right)\exp\left(iv\sigma_{2}\right),
\end{equation}

\noindent with $u=\frac{1}{2}(t+\varphi),v=\frac{1}{2}(t-\varphi)$,
the spectral flow amounts to shifts $u\rightarrow u+w\xi/2,v\rightarrow v+\bar{w}\bar{\xi}/2.$
The coordinates $t,\varphi,\rho$ are global coordinates on AdS\textsubscript{3}
giving the metric $ds^{2}=-\cosh^{2}\rho\, dt^{2}+d\rho^{2}+\sinh^{2}\rho\, d\varphi^{2}.$
Mathematically, the flow transformations (\ref{eq:flow}) are actions
of $\widehat{\textrm{SL}}\textrm{(2,R)}\times\widehat{\textrm{SL}}\textrm{(2,R)}$
loop group elements not continuously connected to the identity. This
is the reason that the \emph{compact} time-like direction $\sigma_{2}$
in SL(2,R) appears singled out in (\ref{eq:flow}). Winding in this
direction cannot be undone. The physical anti-de Sitter spacetime
does not contain closed time time-like curves and is the universal
cover of SL(2,R), but a subset of the flow transformations still act
properly on AdS\textsubscript{3} WZW theory, because the spectral
flow of a WZW solution periodic in $\sigma$ will remain periodic
with the same period if $\bar{w}=w$ is an integer. 

The classical string solutions are coarsely characterized by $w$
and the conjugacy class of their monodromy $M$, which is defined
by periodicity properties. For a periodic solution (\ref{eq:solution})
it is enough for $g\rightarrow gM$ and $\bar{g}\rightarrow M^{-1}\bar{g}$
under the periodicity, and one may check that all $M$ in the same
conjugacy class have the same effect. There can be classical periodic
string solutions for any integer $w$, and for all conjugacy classes,
which can divided into elliptic, parabolic and hyperbolic, distinguished
by the value (sign) of the two SL(2,R)$\times$SL(2,R) Casimirs (which
are equal by correspondence with the point-particle limit). It also
makes special sense to consider the values of the time-like Cartan
generators. Together they give the energy and spin of the strings. 

There are in principle six qualitatively different cases, depending
on the kind of conjugacy class and on whether $w=0$ or not. For $w=0$
the Virasoro constraints allow no solutions with hyperbolic conjugacy,
while parabolic conjugacy is possible for solution without waves on
the string, and any waves on the string are accompanied by a elliptic
conjugacy (corresponding to massive strings in Minkowski space). For
$w\neq0$ the Virasoro constraints relate the total energy and angular
momentum of the solutions to the waves along the string. The elliptic
$w\neq0$ solutions are like positive energy bound states of strings,
hyperbolic $w\neq0$ solutions are the long strings of \citep{MaldacenaMichelsonStrominger1999,SeibergWitten1999}
which expand to arbitrary sizes for asymptotic early and late times,
and parabolic $w\neq0$ solutions are marginally stable and do not
expand.

\paragraph*{Gauge fixing the string.}

The crucial observation in this letter is that physical string solutions
and physical gravity solutions can be transformed into each other
locally. The description of gravity above is complete, in the sense
that the boundary conditions that have been imposed ensure that only
physical degrees of freedom are described by the reduced WZNW model,
but the description of strings is not completely physical in the same
way.  For the AdS\textsubscript{3} string, a price has to be payed
for completely gauge fixing to physical degrees of freedom. Just as
manifest Lorentz invariance is lost in the light cone gauge for Minkowski
strings, here we expect to lose manifest AdS symmetry.

The gauge symmetry of the string is reparametrization symmetry. It
is partially fixed by choosing the conformal gauge (a conformally
flat 2d metric) with the residual gauge transformations, conformal
transformations, preserving conformal flatness. To fix further, a
relation between spacetime and world-sheet is typically used, defining
the physical meaning of the gauge fixed world-sheet coordinates.
The conditions $\frac{d}{d\tau}\left(X^{0}+X^{D-1}\right)=\mathrm{const}$
and $\frac{d}{d\sigma}\left(P^{0}+P^{D-1}\right)=\mathrm{const}$
on coordinates and momentum densities define $\tau$ and $\sigma$.
In AdS\textsubscript{3} I instead use equations (\ref{eq:constraints})
which have a similar structure, to fix the conformal gauge freedom.
Just as in Minkowski space, current conservation ensures that the
gauge fixing is consistent over time. Another required property of
a gauge fixing is that it should actually fix the gauge: the constraint
should be possible to solve, uniquely for a complete gauge fixing.
In the Hamiltonian formalism this is expressed by the requirement
that the gauge constraint and the gauge fixing should have non-degenerate
Poisson bracket (allowing the construction of a Dirac bracket). So
we need the bracket of the constraints (\ref{eq:Virasoro}) and (\ref{eq:constraints}).
The Fourier components of the energy-momentum tensor are conventionally
denoted by $L$ and we get the component equations
\begin{equation}
\begin{array}{c}
\left\{ L_{m},J_{n}^{+}\right\} =nJ_{m+n}^{+}=n\kappa\delta_{m,-n},\\
\left\{ \bar{L}_{m},\bar{J}_{n}^{-}\right\} =n\bar{J}_{m+n}^{-}=-n\kappa\delta_{m,-n},
\end{array}
\end{equation}

\noindent where the constraints (\ref{eq:constraints}) have been
used in the last steps. The gauge fixing is complete except for the
zero modes, $ $$J_{0}^{+}$ and $\bar{J}_{0}^{-}$. We might need
extra zero mode conditions to fix $L_{0}$ and $\bar{L}_{0},$ but
this is actually reminiscent of the special role of the $ $$L_{0}-\bar{L}_{0}$
constraint in the Minkowski light cone gauge.

\paragraph*{The relation between strings and gravity.}

In gravity we have the light-like $J\left[E_{\pm}\right]=\pm\kappa$
and the space-like constraints $ $$J\left[H\right]=\bar{J}\left[H\right]=0$,
and we have now gauge fixed to the same light-like constraint for
a string, but with the space-like constraint replaced by the Virasoro
conditions (\ref{eq:Virasoro}), $T=\bar{T}=0.$ To get another perspective
on the two differently constrained systems I propose to invert the
picture: Regard $J\left[E_{\pm}\right]=\pm\kappa$ as the fundamental
constraint and $ $$J\left[H\right]=\bar{J}\left[H\right]=0$ or $T=\bar{T}=0$
as \emph{alternative gauge fixings. }This means that (at least locally)
asymptotic AdS\textsubscript{3 }gravity solutions can be transformed
to AdS\textsubscript{3} string solutions and conversely. However,
because the complete set of constraints differs, care has to be taken
in extracting the observables (the Dirac brackets are different).
 Another potentially subtle point is that the $ $$J\left[H\right]=\bar{J}\left[H\right]=0$
constraints fix the freedom in going from a Liouville solution to
a WZNW solution algebraically, while the Virasoro constraints only
fix a derivative, potentially leaving more solutions, depending on
the boundary conditions.

Finding string solutions from gravity solutions and conversely involves
no other problems of principle. I have studied a few simple cases
to check how the mapping works in practice. It is of some interest
to find out what the curious winding solutions of AdS\textsubscript{3}
strings correspond to viewed as gravity solutions. It turns out that
string winding solutions ($w\neq0$) generically give rise to Liouville stress
tensors with localized singularities in $\tau-\sigma$ and $\tau+\sigma$,
which in the gravity solutions of \citep{Banados1998} translates
to spacetime coordinates $t-\varphi$ and $t+\varphi$. This could
explain why such solutions, to my knowledge, have not been considered.
But even if they have singularities in the gauge of \citep{Banados1998},
a typical asymptotic point will not be different from other asymptotic
points of solutions obeying the Brown-Henneaux boundary conditions
globally. Furthermore, the fact that winding solutions played a crucial
role in the solution of AdS\textsubscript{3} string theory suggests
an important role for the corresponding gravity solutions.

\paragraph*{Conclusion.}

In three dimensions string solutions and gravity solutions are related
by a map similar to a gauge transformation. The existence of this
map suggests an alternative picture of gravity. At least in the case
studied here, gravity is a theory of a (hyper-)surface in a symmetric
space. The shape of this hyper-surface encodes the possible spacetime
solutions. More concretely, the AdS\textsubscript{3} case studied
here is related to a specific string theory with a reasonably well
understood quantum theory. An attempt is under way to connect that
quantum theory to quantum gravity in analogy to the the classical
discussion in this letter.
\begin{acknowledgments}
I wish to thank Igor Klebanov and the Princeton Physics Department,
and the Galilieo Galilei Institute for Theoretical Physics and INFN,
for their hospitality and partial support in important phases of this
work. I have also benefited greatly from discussions with Ed Witten and
Massimo Porrati.
\end{acknowledgments}
\bibliography{/Users/bosundborg/Documents/Bibliography}

\end{document}